\documentclass[12pt]{article}
\begin{document}

\title{{\bf Hyper-Entropic Gravitational Fireballs (Grireballs) with Firewalls}}

\author{
Don N. Page
\thanks{Internet address:
profdonpage@gmail.com}
\\
Department of Physics\\
4-183 CCIS\\
University of Alberta\\
Edmonton, Alberta T6G 2E1\\
Canada
}

\date{2012 November 28; slightly revised 2013 March 27}

\maketitle
\large
\begin{abstract}
\baselineskip 25 pt

Recently there has been much discussion as to whether old black holes have firewalls at their surfaces that would destroy infalling observers.  Though I suspect that a proper handling of nonlocality in quantum gravity may show that firewalls do not exist, it is interesting to consider an extension of the firewall idea to what seems to be the logically possible concept of hyper-entropic gravitational hot objects (gravitational fireballs or {\it grireballs} for short) that have more entropy than ordinary black holes of the same mass.  Here some properties of such grireballs are discussed under various assumptions, such as assuming that their radii and entropies both go as powers of their masses as the one independent parameter, or assuming that their radii depend on both their masses and their entropies as two independent parameters. 

\end{abstract}

\normalsize

\baselineskip 23 pt

\newpage

Recently Almheiri, Marolf, Polchinski, and Sully \cite{AMPS} have given a provocative argument that suggests that an ``infalling observer burns up at the horizon'' of a sufficiently old black hole, so that the horizon becomes what they called a ``firewall.''  This paper has elicited a large number of responses, some of which support the firewall idea \cite{Susskind:2012rm, Susskind:2012uw, Bousso:2012as, Giveon:2012kp}, others of which raise skepticism about it \cite{Nomura:2012sw, Mathur:2012jk, Chowdhury:2012vd, Banks:2012nn, Ori:2012jx, Brustein:2012jn, Hossenfelder:2012mr, Nomura:2012cx, Avery:2012tf, Larjo:2012jt, Rama:2012fm}, and yet others seem more agnostic \cite{Bena:2012zi, Hwang:2012nn}.  

(Samuel Braunstein has informed me that he and Pirandola and \.{Z}yczkowski had written a paper \cite{Braunstein:2009my} that had earlier derived the identical phenomenon which they called ``energetic curtains,'' but that paper makes rather different assumptions and comes to the conclusion that ``the information entering a black hole \ldots\ is only decoded into the outgoing radiation very late in the evaporation,'' which is in conflict with the conclusions of Hayden and Preskill \cite{Hayden:2007cs} that ``Information deposited prior to the half-way point remains concealed until the half-way point, and then emerges quickly.'')

I do not yet see a clear resolution of the firewall puzzle, though my tentative intuition is that when the nonlocality of gravity is properly taken into account, a unitary theory of quantum gravity will not give firewalls for typical black holes, no matter how old they are.  Some partial indications in that direction are given in \cite{Nomura:2012sw, Banks:2012nn, Brustein:2012jn, Hossenfelder:2012mr, Nomura:2012cx}, though I do not think they or any other current papers have yet given a completely proper account.  We simply do not know enough about quantum gravity at present to be able to describe quantum black holes correctly.

Here I wish to suggest a possibility that, although I personally think is unlikely to be true, seems to be at least logically consistent with what little we do know for certain about quantum gravity:  the existence of quasi-stable gravitational objects that have even larger entropies (or number of microstates up to some energy) than black holes.  I shall assume that such objects can be formed individually in a unitary theory (e.g., not only by pair production, such as the way an electron-positron pair can be produced from an electromagnetic field when there are no initial charged particles present, even though one cannot produce an individual electron from such an initial condition), so that such objects can also decay individually (and not only by pair annihilation).  Therefore, in suitable quantum states that I shall assume are generic, such an object in quasi-stable thermal equilibrium (i.e., ignoring the decay from radiation off its surface) can be characterized by a temperature and can be considered to be a ``hot'' object (though I do not mean to assume that the temperature is necessarily high by any particular standard).  As hot gravitating objects, I shall call them gravitational fireballs, or {\it grireballs} for short.

Here I shall use Planck units $\hbar = c = G = k \equiv k_{\mathrm{Boltzmann}} =
4\pi\epsilon_0 = 1$.  For simplicity, I shall here focus on grireballs with
fairly definite mass $M \gg 1$ and essentially zero angular momentum ($J \ll
M^2$) and assume that they are macroscopically spherically symmetric.  I shall
further assume that when their temperatures are sufficiently low that for times
sufficiently short that one can ignore the thermal radiation from the
grireballs, there are quantum states in asymptotically flat spacetime with
negligible quantum excitations (e.g., particles and radiation) surrounding the
grireballs, so that they may be considered to have fairly-definite radii $R >
2M$, outside of which the quantum state is locally close to the Boulware vacuum
\cite{Boul} of the Schwarzschild metric with mass $M$.  That is, I shall focus
on macroscopically distinct quantum states in which, to a good approximation,
outside $R$ the gravitational field is close to that of the Schwarzschild
metric, and the expectation values of the stress-energy tensor is much less than
the orthonormal Riemann tensor component $\sim M/R^3$ in the Schwarzschild
metric near the surface of the grireball (not that the small nonzero expectation
values of the stress-energy tensor are being assumed to be relatively closer to
the Boulware vacuum values than, say, to the Unruh \cite{Unruh} or
Hartle-Hawking \cite{HH} vacuum, but only that they do not have a large effect on
the geometry for the region outside but moderately close to the surface of the
grireball, $r \sim R$, but not making such a restriction for $r \gg R$, at which
the cumulative effect of the radiation could cause the Misner-Sharp mass
\cite{MS} $m(r)$ to have significant departures from $M = m(R)$, the mass of the
grireball itself, if the total energy $E = m(\infty)$ of the quantum state is
significantly larger than $M$ from the dilute radiation that I am
allowing to surround the grireball).

I shall be agnostic as to what is ``inside'' the grireball.  If it is not enormously larger than a black hole of the same mass and yet has more entropy $S(M)$ for a given mass $M$ than an ordinary Schwarzschild black hole, which has $S = \pi R^2 = 4 \pi M^2$, presumably the grireball must not be made up of ordinary matter in a spacetime metric obeying some approximation to the classical or semiclassical Einstein equations.  I shall not assume that there is necessarily any spacetime at all inside the fireball, but only that from the outside, in the thermodynamic approximation for a generic state in which that is valid, the grireball appears to have a surface of radius $R$ and temperature that would redshift to the value $T(M) = dM/dS(M)$ at spatial infinity ($r=\infty$) if there were negligible surrounding radiation to distort the metric significantly away from the Schwarzschild form, so that $E-M \ll M$.  In more detail, the temperature $T(M)$ is the temperature with respect to an approximate (e.g., ignoring the time dependence from the evaporation) timelike Killing vector $\partial/\partial t$ normalized so that $g_{tt}g_{rr} = -1$ in the approximate Schwarzschild metric just outside the grireball.  That is, the time coordinate is normalized so that its imaginary periodicity in the thermal Green functions is equal to $\sqrt{g_{rr}}$ divided by the local temperature measured by a static observer at $r \sim R$, neglecting the assumed tiny difference between $m(r)$ and $M = m(r)$ for $r \sim R$.

Essentially, I am first assuming that grireballs are macroscopically spherically symmetric and have radii $R$ and entropies $S$ that are functions of their masses $M$.  That is, I shall take them to form a one-parameter family, with $R = R(M)$ and $S = S(M)$.  However, near the end of this paper I shall consider two-parameter families of grireballs, with $R = R(M,S)$.

Let me list ten possible assumptions (each labeled below by a short name) for this macroscopically one-parameter family of spherically symmetric hyper-entropic gravitational fireballs, the first three of which I shall take to be part of the definition of such grireballs, and the other seven of which may be taken to be possible additional assumptions of various degrees of plausibility or applicability:

\begin{enumerate}

\item Macroscopically spherically symmetric, with $R = R(M)$ and $S = S(M)$.

\item Hyper-entropic:  $S > 4\pi M^2$

\item Visible:  $R > 2M$

\item Large:  $R \geq 3M$

\item Hawking:  $S \leq \pi R^2$

\item Bekenstein:  $S \leq 2\pi R M$

\item Geometric-optics:  $\frac{\lambda_{\mathrm{thermal}}}{R}
                          = \frac{2\pi}{R T} \ll 1$

\item Blackbody:  surface totally absorbing

\item Cool:  negligible massive particles emitted

\item Slowly evaporating:  $-dR/dt < 1$

\end{enumerate}

One can see that Assumptions 2 and 5 imply 3, as do Assumptions 2 and 6, but Assumptions 2 and 3 do not imply either 5 or 6 (or 4, 7, 8, 9, or 10 for that matter).  Also, Assumptions 3 and 6 imply 5, but 3 and 5 do not imply 6.  Since I am always assuming 1, 2, and 3 as part of the definition of grireballs, Bekenstein grireballs (those obeying Bekenstein's conjectured entropy bound \cite{Bek}) are necessarily also Hawking grireballs (those with entropy not more than the Hawking entropy $A/4$ \cite{Haw} of a black hole of the same area), but Hawking grireballs are not necessarily Bekenstein grireballs.

The local temperature at the surface of a (visible, so $R > 2M$) grireball is $T_{\mathrm {surface}} = T (1 - 2M/R)^{-1/2}$, where $T = dM/dS$.  In the region near but outside the grireball surface, where I am assuming that one can neglect the effect of the stress-energy tensor, the Misner-Sharp mass is nearly constant, $m(r) \approx M$, and the normalization above for the time coordinate $t$ makes $-g_{tt} \approx (g_{rr})^{-1} \approx (1 - 2M/r)$ and the locally measured temperature $T_{\mathrm{local}} = T(-g_{tt})^{-1/2} \approx T (1 - 2M/r)^{-1/2}$.

Large, geometric-optics, blackbody, cool grireballs emit essentially only massless radiation that then very nearly all escapes to infinity.  In the frame of a local static observer at $r = R$, the blackbody emission luminosity is $L_{\mathrm{surface}} = \sigma T_{\mathrm{surface}}^4 4 \pi R^2$, where $\sigma = a/4$ is the Stefan-Boltzmann constant ($\pi^2/60$ for photons or for gravitons separately, or $\pi^2/30$ for both) and here (but often not below) $a$ is the radiation constant ($\pi^2/15$ for photons or for gravitons separately, or $2\pi^2/15$ for both) for the radiation assumed to be emitted, and $4 \pi R^2$ is the area of the surface of the grireball.  However, relative to the $\partial/\partial t$ Killing vector field normalized so that $g_{tt}g_{rr} = -1$ in the approximate Schwarzschild metric just outside the grireball, there is one redshift factor for the energy and another for the number rate, giving a normalized luminosity of 
\begin{equation}
L  = \sigma T_{\mathrm{surface}}^4 4 \pi R^2 (1 - 2M/R)
   = \frac{4 \pi R^2 \sigma T^4}{(1 - 2M/R)}
   = \frac{(2 \pi^3/15) R^2 T^4}{(1 - 2M/R)},
\label{luminosity}
\end{equation}
where for the last expression I have used the Stefan-Boltzmann constant $\sigma
= \pi^2/30$ for the assumed blackbody emission for both photons and gravitons. 
The enhancement factor $(1 - 2M/R)^{-1}$ can be interpreted as the enhancement
factor, over $\pi R^2$, for the cross section of the surface as seen by a
distant observer.  

If the grireball is not ``large'' but instead has $R(M) < 3M$, and yet with the other Assumptions 7, 8, and 9 still holding, only a fraction of the emission directions from the emission surface will actually propagate outward through the centrifugal barrier at $r = 3M$.  From far away, the surface appears to have an effective cross section the same as if it were at $R = 3M$, so the normalized luminosity then is $L = 12 \pi R^2 \sigma T^4$.

If one gives up Assumption 7 of geometric optics, the luminosity would be decreased, perhaps by a factor approximately proportional to $(RT)^2$ when this quantity is large.  If one gives up Assumption 8 of a blackbody surface, the thermal luminosity would also be reduced by the factor of the absorptivity.  On the other hand, if Assumption 9 of coolness is violated, then the grireball can thermally emit massive particles, which would increase the luminosity.

Since grireballs are highly hypothetical, in principle they could have any of a wide range of possible functional forms for their radii $R(M)$ and entropies $S(M)$ as functions of their masses $M$.  Or, later I shall consider the possibility that their properties depend on two parameters, say $M$ and $S$.  However, here for simplicity in the one-parameter case I shall just consider simple power-law relations, 
\begin{equation}
R = a M^{\alpha},\ \ S = b R^{\beta} = a^\beta b M^{\alpha\beta} = c M^\gamma,
\label{radius-entropy}
\end{equation}
with $c = a^\beta b$, $\gamma = \alpha\beta$, and with $a$ no longer the radiation constant.

One can now get the temperature of the grireball as
\begin{equation}
T = \frac{dM}{dS} = \frac{1}{a^\beta b \alpha\beta M^{\alpha\beta - 1}}
                 = \frac{1}{c \gamma M^{\gamma - 1}}.
\label{temperature}
\end{equation}

Then the luminosity from large, geometric-optics, blackbody, cool grireballs is
\begin{equation}
L  = \frac{(2 \pi^3/15)}
{(1-2M^{1-\alpha}/a)a^{4\beta-2}(b\alpha\beta)^4 M^{4\alpha\beta-2\alpha-4}}
\label{lum}
\end{equation}
Setting $dM/dt = -L$ then gives the lifetime for a large, geometric-optics, blackbody, cool grireball of mass $M \ll R = aM^\alpha$ to decay as
\begin{eqnarray}
t_{\mathrm{lifetime}}  
&=& \frac{a^{4\beta-2}(b\alpha\beta)^4 M^{4\alpha\beta-2\alpha-3}}
       {(2 \pi^3/15)(4\alpha\beta-2\alpha-3)} \nonumber\\
&=& \frac{15\gamma^4 c^4}{2\pi^3a^2}
     \frac{M^{4\gamma-2\alpha-3}}{4\gamma-2\alpha-3}.
\label{lifetime}
\end{eqnarray}

We can also calculate the time $t_\ast$ for a large, geometric-optics, blackbody, cool grireball to emit radiation with entropy equal to that left in the grireball.  (That is the time at which one might expect that the quantum state would give nearly maximal entanglement between the radiation and what remains of the radiating object \cite{Page:1993df, Page:1993wv}.)  The entropy of the massless radiation in the geometric-optics blackbody limit is 4/3 its energy divided by its temperature, whereas the entropy decrease in the grireball during a relatively small energy loss is that energy loss divided by the temperature, so the emission into empty space increases the total entropy by a factor of $f = 4/3$.  That is, if $S_0 = cM_0^\gamma$ is the initial entropy of the grireball, after the grireball has lost energy and entropy down to entropy $S$, the entropy in the radiation is $f(S_0-S)$.  Setting these two equal gives the entropy then remaining in the grireball at $t_\ast$ as $S = S_\ast = [f/(f+1)]S_0 = (4/7)S_0$.  Abbreviating the grireball lifetime given by Eq.\ (\ref{lifetime}) as $t_f \equiv t_{\mathrm{lifetime}}$, the evolution of a large, geometric-optics, blackbody, cool grireball with initial mass $M_0$, initial radius $R_0 = aR^\alpha$, and initial entropy $S_0 = cM^\gamma$ at time $t=0$ gives
\begin{equation}
M = M_0\left(1 - \frac{t}{t_f}\right)^{\frac{1}{4\gamma-2\alpha-3}},\ \ 
R = R_0\left(1 - \frac{t}{t_f}\right)^{\frac{\alpha}{4\gamma-2\alpha-3}},\ \ 
S = S_0\left(1 - \frac{t}{t_f}\right)^{\frac{\gamma}{4\gamma-2\alpha-3}}.
\label{griballevolution}
\end{equation}
Therefore, when the entropy radiated equals the entropy remaining in the grireball, one gets
\begin{equation}
t_\ast = 
t_f\left[1-\left(\frac{4}{7}\right)^{\frac{4\gamma-2\alpha-3}{\gamma}}\right],
\ \ 
M_\ast = \left(\frac{4}{7}\right)^\frac{1}{\gamma} M_0,\ \ 
R_\ast = \left(\frac{4}{7}\right)^\frac{\alpha}{\gamma} R_0,\ \ 
S_\ast = \frac{4}{7} S_0.
\label{equalvalues}
\end{equation}

Now let us see what restrictions there are on the exponents $\alpha$ and $\beta$ (and hence also on $\gamma$) from the ten possible grireball assumptions listed above in the limit that the mass $M$ is taken to infinity:

\begin{enumerate}

\item Macroscopically spherically symmetric, with $R = R(M)$ and $S = S(M)$.

\item Hyper-entropic:  $S > 4\pi M^2 \Rightarrow \alpha\beta = \gamma \geq 2$

\item Visible:  $R > 2M \Rightarrow \alpha \geq 1$

\item Large:  $R \geq 3M \Rightarrow \alpha \geq 1$

\item Hawking:  $S \leq \pi R^2 \Rightarrow \beta = \gamma/\alpha \leq 2$

\item Bekenstein:  $S \leq 2\pi R M
                   \Rightarrow \alpha\beta = \gamma \leq \alpha+1$

\item Geometric-optics:  $\frac{\lambda_{\mathrm{thermal}}}{R}
                          = \frac{2\pi}{R T} \ll 1
			   \Rightarrow \alpha\beta = \gamma \leq \alpha+1$

\item Blackbody:  surface totally absorbing

\item Cool:  negligible massive particles emitted 
             $\Rightarrow \alpha\beta = \gamma \geq 1$

\item Slowly evaporating:  $-dR/dt < 1 
             \Rightarrow \alpha\beta = \gamma \geq (3/4)(\alpha+1)$

\end{enumerate}

The implication from the last assumption (slowly evaporating, by which I mean that the inward velocity of the grireball surface is not faster than the speed of light, at least if $R \gg 2M$) also assumes that one has a large, geometric-optics, blackbody, cool grireball.  However, one should note that there is not really any causality restriction on the inward velocity of the grireball surface, since it need not carry information inward but can just be a mathematically defined surface chosen to demarcate what is considered to be the outer edge of the grireball.  For example, a sphere of electrons and positrons could annihilate in such a way that the surface that marks the outer boundary of the region where the electron-positron density is still high could move inward at an arbitrarily high velocity.

Although one could consider grireballs (defined to obey Assumptions 1-3) that violate any of the Assumptions 4-10, there is a range in the two-parameter set of exponents for the power-law relations $R = a M^{\alpha}$ and $S = c M^\gamma$ (or $S = b R^{\beta}$ with $\beta = \gamma/\alpha$ and $b = c/a^{\beta}$) that for large enough $M$ gives grireballs that obey all ten Assumptions:
\begin{equation}
1 \leq \gamma - 1 \leq \alpha \leq \frac{4}{3}\gamma - 1.
\label{exponents}
\end{equation}
The last inequality is only needed for the slowly-evaporating assumption, so one might consider abandoning it, but here for simplicity I shall keep them all unless otherwise stated.  

One can also derive expressions for the minimal mass of grireballs to obey the assumptions for given exponents $\alpha$ and $\gamma$ and for given coefficients $a$ and $c$, but I shall not do that here.  However, I can note that if the coefficients obey the following sequence of inequalities, then (except for the coolness Assumption 9, which requires $M \gg (c\gamma m)^{-1/(\gamma - 1)}$ where $m$ is the mass of the lightest massive particle, such as a neutrino or an electron with $m \approx 4\times 10^{-23}$ Planck units \cite{RPP}), all of the assumptions would hold for all grireballs of mass greater than the Planck mass if the following sequence of inequalities held:
\begin{equation}
8\pi^2\gamma+3 \ll 2\pi c\gamma+3 \ll a \ll 
\frac{2}{\pi}\left(\frac{\gamma^4}{\alpha}\right)^{1/3} c^{4/3}.
\label{coefficients}
\end{equation}
On the other hand, since one would normally be considering grireballs much more massive than the Planck mass, it would then be sufficient for all the assumptions (except the coolness Assumption 9, which requires $M \gg m^{-1} \sim 10^{22}$ or perhaps a few orders of magnitude greater from the mass of the lightest neutrino if it is a few orders of magnitude smaller than the mass of an electron, grireball mass $M$ very much more massive than the Planck mass, though much less than a solar mass $M_\odot \approx 10^{38}$ in Planck units \cite{RPP} if the lightest neutrino mass is much greater than $10^{-38}$ Planck units or than about $10^{-10}$ eV) if the exponents obeyed the inequalities (\ref{exponents}) as strict equalities (i.e., $1 < \gamma - 1 < \alpha < (4/3)\gamma - 1$).

It is interesting that the lifetime of a large, geometric-optics, blackbody, cool, slowly evaporating grireball for very large mass $M$, which by Eq.\ (\ref{lifetime}) is proportional to $M^{4\gamma-2\alpha-3}$, can be either be shorter or longer than the lifetime of a black hole of the same mass, which is proportional to $M^3$, and yet be consistent with all of the other assumptions if $2 < \gamma < 3$.  However, if $\gamma > 3$, then such a grireball necessarily has a longer lifetime than a black hole of the same mass in the limit that the mass is taken to infinity.  In particular, $\alpha > 2\gamma - 3$ makes $4\gamma-2\alpha-3 < 3$ and hence gives the large-mass grireball a shorter lifetime than a black hole of the same mass, but then $2\gamma - 3 < \alpha < (4/3)\gamma - 1$ requires $\gamma < 3$ as well as the hyper-entropic assumption $\gamma \geq 2$.  On the other hand, $\alpha < 2\gamma - 3$ makes the large, geometric-optics, blackbody, cool, slowly evaporating grireball lifetime go as a higher power of the mass $M$ than it does for a black hole, and this range for $\alpha$ is consistent with the other assumptions for all super-entropic values of $\gamma$, $\gamma \geq 2$.

Let us compare a grireball with a black hole, which has $\alpha = 1$, $\beta = 2$, $\gamma = 2$, $a = 2$, $b = \pi$, and $c = 4\pi$.  A black hole does not qualify as a grireball, since it is not ``hyper-entropic'' ($S > 4\pi M^2$) and is not ``visible'' ($R > 2M$), and it is also not ``large'' ($R \geq 3M$) or ``geometric-optics'' ($2\pi/(RT) \ll 1$) but rather has $2\pi/(RT) = 8\pi^2 \gg 1$ and hence emits radiation with a characteristic thermal wavelength $\lambda_{\mathrm{thermal}} = 2\pi/T = 16\pi^2 M = 8\pi^2 R$ many times longer than the black hole radius $R = 2M$.

Numerical calculations \cite{Page1,Page2,Page3,Page4} have shown that a nonrotating black hole large enough (e.g., solar mass or greater, assuming that the lightest neutrino has a mass at least about $10^{-10}$ eV) to emit essentially only photons and gravitons loses mass and entropy, and emits radiation entropy, at the rates
 \begin{equation}
 -\frac{dM}{dt} \approx \frac{3.7475\times 10^{-5}}{M^2},
 \label{massloss}
 \end{equation}
 \begin{equation}
 -\frac{dS_{\mathrm{bh}}}{dt} \approx \frac{0.9418\times 10^{-3}}{M},
 \label{entropyloss}
 \end{equation}
 \begin{equation}
 \frac{dS_{\mathrm{rad}}}{dt} \approx \frac{1.3984\times 10^{-3}}{M}.
 \label{entropyradiated}
 \end{equation}
Therefore, a large nonrotating black hole emits photon plus graviton radiation entropy of $f_{\mathrm{bh}} = -dS_{\mathrm{rad}}/dS_{\mathrm{bh}} \approx 1.4847$ times as much as the entropy decrease of the black hole, a rather larger fraction than the $f = 4/3$ for a blackbody emitting purely massless thermal radiation of wavelength much shorter than its size, such as a large, geometric-optics, blackbody, cool grireball \cite{Page:1983ug,Page:2013dx}.

From these numbers, one can calculate that the lifetime of a large black hole is
\begin{equation}
t_{\mathrm{bh}}\approx 8895\, M_0^3
 \approx 6.7836\times 10^{117}\left(\frac{M_0}{M_\odot}\right)^3
 \approx 1.1589\times 10^{67}\left(\frac{M_0}{M_\odot}\right)^3 \mathrm{years},
\label{holelife}
\end{equation}
and that the time to emit radiation with an entropy equal to that remaining in the black hole is
\begin{equation}
t_{\mathrm{bh}\ast} = 
\left[1-\left(\frac{f_{\mathrm{bh}}}{1+f_{\mathrm{bh}}}\right)^{3/2}\right]
t_{\mathrm{bh}}
\approx 0.5381\, t_{\mathrm{bh}}.
\label{holeentanglementtime}
\end{equation}
The numerical values of this time are then
\begin{equation}
t_{\mathrm{bh}\ast}\approx 4786\, M_0^3
 \approx 3.6503\times 10^{117}\left(\frac{M_0}{M_\odot}\right)^3
 \approx 0.6236\times 10^{67}\left(\frac{M_0}{M_\odot}\right)^3 \mathrm{years},
\label{holeentanglementtimenumerical}
\end{equation}

Furthermore, the values of the black hole mass and entropy at this time when the entropy equals the entropy in the radiation are
\begin{equation}
M_{\mathrm{bh}\ast} = \left(\frac{f_{\mathrm{bh}}}{1+f_{\mathrm{bh}}}\right)^{1/2} M_0
\approx 0.7730\, M_0,
\label{massatequality}
\end{equation}
\begin{equation}
S_{\mathrm{bh}\ast} = \left(\frac{f_{\mathrm{bh}}}{1+f_{\mathrm{bh}}}\right) S_0
\approx 0.5975\, S_0.
\label{entropyatequality}
\end{equation}

If we take a black hole of a solar mass, $M = M_\odot \approx 10^{38}$, such a black hole has a radius I shall call $R_\odot = 2M_\odot \approx 3$ km \cite{RPP} and an entropy $S_\odot = 4\pi M_\odot^2 \approx 10^{77}$, which is approximately the same as the entropy contained in the cosmic microwave background photons at a temperature of $T \approx 2.7$ K $\approx 2\times 10^{-32}$ in Planck units \cite{RPP} in a volume of 2.4 cubic megaparsecs, which is the volume of a sphere of radius approximately 800 kiloparsecs, which if centered on our Solar System would just about encompass the Andromeda Galaxy at a mean distance of about 780 kpc and a diameter of about 43 kpc.  The mass-energy of these cosmic microwave background photons of the same entropy of a solar mass black hole would be $(3\pi/2 M_\odot)(kT/\hbar c)R_\odot \approx 1.7\times 10^7 M_\odot$, where I am using $R_\odot = 2GM_\odot/c^2 \approx 2953$ meters to be the Schwarzschild radius of a black hole the mass of the sun (and {\it not} the radius of the sun itself), $T \approx 2.7$ K as the temperature of the microwave background photons, and $k/\hbar c \approx 437$ m$^{-1}$ K$^{-1}$ \cite{RPP}, so $kT/\hbar c \approx 1200$ m$^{-1}$ and $(kT/\hbar c)R_\odot \approx 3.5\times 10^6$, which is $1/(4\pi)$ times the ratio of the cosmic microwave temperature $T$ to the solar-mass black hole Hawking temperature that I shall call $T_\odot = (\hbar c/k)/(4\pi R_\odot) = 6.17\times 10^{-8}$, giving $T/T_\odot \approx 44\,000\,000$. 

Except for the fact that it does not have a natural edge at radius $R(M)$ as I am assuming that a grireball does, a not-too-large volume of thermal radiation at the present cosmic microwave background temperature $T$ would be like a grireball in being hyper-entropic, having more entropy than that of a black hole of the same mass, $S = 4\pi M^2$.  For example, in flat space with now reverting to using $a = \pi^2/15$ for the radiation constant for photons, the mass-energy of electromagnetic radiation at temperature $T$ in volume $V$ is $M_{\mathrm{rad}} = a T^4 V$, and the entropy is $S_{\mathrm{rad}} = (4/3) a T^3 V$, so
\begin{equation}
\frac{S_{\mathrm{rad}}}{4\pi M_{\mathrm{rad}}^2} = \frac{1}{3\pi a T^5 V}. 
\label{radiation entropy}
\end{equation}

Therefore, thermal electromagnetic radiation in a volume $V < (3\pi a T^5)^{-1} = (0.2\pi^3 T^5)^{-1}$ would be hyper-entropic in comparison with a black hole of the same mass-energy.  For the cosmic microwave background radiation temperature, the maximal volume for which $S \geq 4\pi M^2$ would be $V = (0.2\pi^3 T^5)^{-1} \approx 6\times 10^{157} \approx 9000$ cubic parsecs, which is the volume of a sphere of a radius about 13 parsecs or 40 light years.  Centered on our Solar System, such a sphere would contain roughly a thousand stars.  The mass energy of the cosmic microwave photons in this maximal volume would be $M = (3\pi T)^{-1} \approx 5.5\times 10^{30} \approx 1.2\times 10^{23}$ kilograms, which is about 1.6 times the mass of the Moon.  The entropy of such photons in this maximal volume for the cosmic microwave background to give a hyper-entropic system would be $S_{\mathrm{rad}} = 4/(9\pi T^2) \approx 4\times 10^{62} \approx 4\times 10^{-15} S_\odot$, over 14 orders of magnitude smaller than the entropy $S_\odot = 4 \pi M_\odot^2$ of a solar-mass black hole.

This means that thermal radiation at the cosmic microwave background radiation temperature or higher is hyper-entropic only with respect to black holes much less massive than stars, and in fact less massive than the Earth.  For thermal electromagnetic radiation to be hyper-entropic with respect to a solar-mass black hole, one needs $T < (3\pi M_\odot)^{-1} = (8/3) T_\odot \approx 1.6\times 10^{-7}$ K and $V > (0.2\pi^3 T^5)^{-1} > 3^5 5\pi^2 M_\odot^5 \approx 8\times 10^{193}$, which is the volume of a sphere of radius $r \approx 2.3\times 10^{64} \approx 4.5\times 10^{13}$ light years, which with $c=1$ is approximately 3300 times the present age of the universe that is about 13.75 Gyr or five trillion days or $8.05\times 10^{60}$ Planck times \cite{RPP}.  Therefore, one cannot have ordinary thermal radiation in a region smaller than the observable size of the universe that is hyper-entropic with respect to black holes of solar mass or greater.  If there are hypothetically possible objects that could form within a size not greater than the size of the observable universe (or not greater than the length scale corresponding to the observed value of the apparent cosmological constant) and yet have more entropy for the same mass than stellar-mass black holes, they must be unknown objects such as the hyper-entropic gravitational fireballs (grireballs) postulated in this paper. 

One could also calculate how large a black hole would have the same mass and the same entropy of a ball of thermal electromagnetic radiation in flat spacetime that has a radius equal to the radius of the horizon of de Sitter spacetime with the currently measured value of the cosmological constant, and an appropriate temperature.  For getting a precise nominal value for the cosmological constant that happens to be in agreement with fairly recent measurements, within their experimental error, use a slight simplification of the Mnemonic Universe Model (MUM) given in \cite{Agnesi,abinitio} to now be a spatially flat universe dominated by dust and a cosmological constant, with present age $t_0 = H_0^{-1} =$ five trillion days ($4.32\times 10^{17}$ seconds, or about 13.69 billion Julian years, or about $8.01\times 10^{60}$ Planck times, rather than my previous MUM version of one hundred million years divided by the fine structure constant, which is about 13.7036 billion years or $8.02\times 10^{60}$ Planck times, or a new coincidence that I noted that the age of the universe in Planck units is approximately one-ninth the golden ratio power of the largest prime computed without computers, or $t_0 = (1/9)(2^{127}-1)^{(\sqrt{5}+1)/2} \approx 8.03\times 10^{60}$ Planck times or about 13.71 billion years, or yet another coincidence I just found that the age of the universe is approximately 80 Gyr divided by the cube root of 200, or about 13.68 Gyr).  In comparison, the estimate of the age of the universe from a six-parameter $\Lambda$CDM fit in the most current ``Review of Particle Phyiscs'' \cite{RPP} is $13.75\pm 0.13$ Gyr or $(8.049\pm 0.076)\times 10^{60}$ Planck times, so the differences of the four different approximations above from the mean of this estimate are all less than about 50\% or so of the listed uncertainty of the current estimate of the age of the universe.  

(However, more recent measurements have given values and uncertainties such that the mnemonic formulas are no longer quite within the uncertainties.  For example, the nine-year WMAP results with baryon acoustic oscillations and Hubble-constant priors \cite{Hinshaw:2012fq} give an age of $13.750\pm 0.085$ Gyr or $(8.049\pm 0.050)\times 10^{60}$ Planck times (whose uncertainty does overlap the mnemonic values above) and a Hubble constant of $H_0 = (69.33\pm 0.88)$ km/s/Mpc or $H_0^{-1} = 14.105\pm 0.179$ Gyr or $(8.257\pm 0.105)\times 10^{60}$ Planck times, which is larger than this WMAP measurement of the age by about 1.8 standard deviations.  The even more precise Planck Satellite data combined with WMAP polarization at low multipoles, high-multipole experiments, and baryon acoustic oscillations \cite{Ade:2013xsa} give a universe age of $13.798\pm 0.037$ Gyr, which is over two standard deviations larger than any of the mnemonic values above, and a Hubble constant of $H_0 = (67.80\pm 0.77)$ km/s/Mpc or $H_0^{-1} = 14.423\pm 0.164$ Gyr or $(8.443\pm 0.096)\times 10^{60}$ Planck times, which is larger than this Planck measurement of the age by about 3.8 standard deviations.  Therefore, the mnemonic formulas, while still being useful as fairly good approximations, do not quite fit with the latest more precise WMAP and Planck data.)

The metric for the MUM model is
\begin{equation} 
ds^2 = -dt^2 + \sinh^{4/3}(1.5H_\Lambda t)(dx^2+dy^2+dz^2),
\label{MUM} 
\end{equation} 
where $H_\Lambda = \sqrt{\Lambda/3}$ is the asymptotic value of the
Hubble expansion rate
\begin{equation} 
H = \frac{\dot{a}}{a} = H_\Lambda\coth{(1.5H_\Lambda t)}.
\label{H} 
\end{equation} 

For $t_0 = H_0^{-1}$, we need $H_\Lambda t_0 = \tanh{(1.5H_\Lambda t_0)}$ or
$H_\Lambda t_0 \approx 0.8586$, and then $t_0 =$ five trillion days gives
$H_\Lambda^{-1} \approx 15.94\ \mathrm{Gyr} \approx 5.03\times 10^{17}$
seconds $\approx 9.33\times 10^{60}$ in Planck units, which is the radius $r =
H_\Lambda^{-1}$ of the corresponding de Sitter horizon.  For thermal photons and
gravitons (radiation constant $a = 2\pi^2/15$) in a sphere of this radius in
flat spacetime, volume $V = (4\pi/3)r^3$, the requirement that the entropy
equals that of a black hole of the same mass is $S = 4\pi M^2$ or $(4/3) a T^3 V
= 4 \pi a^2 T^8 V^2$ or $1 = 3 \pi a T^5 V = 4 \pi^2 a r^3 T^5 = (8 \pi^4/15)
r^3 T^5$, which gives $T = [(8 \pi^4/15) r^3]^{-1/5} \approx 1.19\times
10^{-37} \approx 1.68\times 10^{-5}$ K.  The mass of this radiation is then $M
= a T^4 V = (3 \pi T)^{-1} \approx 9\times 10^{35} \approx 0.01
M_\odot$, or a bit over ten times the mass of Jupiter.  This is the mass of the
largest black hole for which one could imagine a sphere of radiation in flat
spacetime having more entropy for the same mass and yet not having a radius
larger than that of the horizon of the de Sitter spacetime to which our universe
seems to be tending toward if the currently observed cosmic acceleration is due
to a truly constant dark energy density or cosmological constant.

That is, we know of no entities within our present universe that for the same mass can have more entropy than a black hole of more than ten or so times the mass of Jupiter.  However, in this paper I am conjecturing that it would not be inconsistent with the known laws of physics that there exist such hyper-entropic entities, such as perhaps gravitational fireballs or grireballs that are approximately spherically symmetric and have a fairly well defined radius $R(M) > 2M$ and entropy $S(M) > 4\pi M^2$, outside of which the spacetime can be approximately given by the Schwarzschild metric.

As a simple example for a putative grireball size and entropy relation that satisfies all the assumptions above for a grireball mass large in Planck units, suppose $a = b = c = 1$, $\alpha = 1.2$, $\beta = 1.75$, and thus $\gamma = 2.1$, so $R = M^{1.2}$ and $S = R^{1.75} = M^{2.1}$.  Therefore, the entropy obeys the Hawking Assumption 5, $S \leq \pi R^2 = \pi M^{2.4}$ (as well as the Bekenstein Assumption 6, $S \leq 2 \pi R M = 2 \pi M^{2.2}$), so the entropy is less than that of a black hole of the same radius, but it is greater than that of a black hole of the same mass.  This particular set of exponents gives a large, geometric-optics, cool, slowly evaporating grireball for very large mass $M$, and if one further assumes that the grireball is also a blackbody for both photons and gravitons, the lifetime would be $[2.5(2.1)^4/\pi^3]M^3 \approx 1.57 M^3$, going as the same power of $M$ as a black hole, but with a smaller coefficient than that for the lifetime of a large Schwarzschild black hole that emits only photons and gravitons, which from the results above \cite{Page1} is about $8895\, M^3$.  Of course, the coefficient 1.57 for the fireball depends on the choice $a = c = 1$, since for different values of these one would get a lifetime of about $1.57 (c^4/a^2) M^3$.

A grireball of the same mass as the sun but with $a = b = c = 1$, $\alpha = 1.2$, $\beta = 1.75$, and $\gamma = 2.1$ would have a radius $R = M_\odot^{1.2} \approx 3.6\times 10^{45} \approx 5.8\times 10^{10}$ meters $\approx 2\times 10^7 (2M_\odot) \approx 0.4$ AU (astronomical units, one AU being approximately 150 million kilometers or about 500 light-seconds), very close to being the same as the semimajor axis of the orbit of Mercury (0.39 AU), and an entropy $S = M_\odot^{2.1} \approx 5\times 10^{79}$, about 500 times the entropy of a solar-mass black hole and hence approximately the entropy of a $22 M_\odot$ black hole, or approximately the entropy of the cosmic microwave background photons in a volume of 1200 cubic megaparsecs, roughly a hundred times the volume of our Local Group of galaxies.

In considering the hypothetical possibility of super-entropic gravitational fireballs, or grireballs, one should check that they do not violate any current observations.  Potentially one of the most dangerous consequences of grireballs (for ruling them out observationally, not for their threatening our existence) would seem to be their possible production by black holes.  One would imagine that black holes should be able to emit grireballs as part of their Hawking emission, and observationally we require that the rates not be so great as to be in conflict with astronomical black hole observations.

Besides geometric factors, the main factor in the emission rate for grireballs would presumably be the exponential of the change in the thermodynamic (i.e., coarse-grained) entropy of the world during the emission of a grireball.  Suppose we consider a nonrotating black hole of energy $E$ that emits a grireball of mass $M$ to become a black hole of energy $E-M$.  Taking $S(M)$ to be the entropy of the grireball, the change in entropy (neglecting phase-space factors for the motion of the emitted grireball) is $\Delta S = S(M) + 4\pi[(E-M)^2 - E^2] = S(M) - 8\pi E M + 4\pi M^2$.  If this change in entropy is very large (in comparison with unity, not in comparison with the original entropy black hole entropy $4\pi E^2$), then there is a danger that grireballs might be emitted rapidly from a black hole.  Na\"{\i}vely it would appear that this danger indeed occurs when one sets $M = E$, since then $\Delta S = S(E) - 4 \pi E^2 > 0$ by the hyper-entropic property assumed for grireballs.

However, if a grireball with mass $M$ equal to the energy $E$ of the original black hole is much larger than the black hole, say with size $R(M) = a M^\alpha \gg 2M = 2E$, the size of the black hole, there presumably would be severe geometric suppression factors for the rate of emission of the large grireball by the much smaller black hole.  For example, with radiation constant $a = 2\pi^2/15$ for photons and gravitons, entropy would be increased by the conversion of a black hole of energy $E$ to thermal photons and gravitons in a box of volume $V > (3\pi)^4 a^{-1} E^5 \approx 1.4\times 10^{192} (E/M_\odot)^5 \approx 2.7\times 10^9 t_0^3 (E/M_\odot)^5 \approx (4\pi/3) (865 t_0)^3 (E/M_\odot)^5 \approx (4\pi/3) (7400\ \mathrm{light\ years})^3 (E/M_\oplus)^5$, where $M_\oplus \approx 0.0044\,\mathrm{m} \approx 2.7\times 10^{32}$ is the mass of the earth \cite{RPP} (and noticing the mnemonic device that a year in Planck units is, within the experimental uncertainty in the Planck time, 0.1\% larger than $10^{50}$ multiplied by the cube root of 200), so that it would be entropically favorable for a black hole of the mass of the earth to convert to radiation in a box larger than about 2\% of the volume of a sphere with radius equal to our distance of about 27\,000 light years from the Galactic center.  However, because this volume is much larger than the size of the black hole, the time for the conversion would be that of the decay of the black hole by Hawking emission, about $3.14\times 10^{50} (E/M_\odot)^3$ years, so there is no danger of this happening for a large black hole within times that we can observe that are limited by the age of the universe.

Therefore, one might suppose that the danger to our observations of apparent astrophysical black holes that seem not to have converted to grireballs (unless they are actually grireballs that happen to be not much larger than black holes of the same mass) might only come from the emission of grireballs smaller than the black hole, or $R(M) < 2E$.  Then $S(M) - \Delta S = 8\pi E M - 4\pi M^2 > 4\pi R(M) M - 4 \pi M^2$.  Now the assumption that the grireball is `visible,' $R(M) > 2M$, gives $4\pi M^2 < 2\pi R(M) M$, so $S(M) - \Delta S > 2\pi R(M) M$.  Then if we assume the Bekenstein property of grireballs, $S \leq 2\pi R M$, we get that $S(M) - \Delta S > S(M)$, or $\Delta S < 0$, so the rate of grireball emission would be suppressed.  (I have ignored the phase space factors for the motion of the emitted grireballs, so in fact they can be emitted without decreasing the entropy, just as other particles can be emitted by Hawking radiation, but the intrinsic entropy of the grireballs will not give a huge enhancement factor if the $\Delta S$ calculated above is not much larger than unity.)

As a result, if there is indeed a sufficient geometric suppression for the emission of grireballs larger than the black hole, then it appears that something like the Bekenstein assumption for them would be sufficient to avoid the danger of black holes quickly (e.g., on timescales less than the age of the universe) turning into grireballs.  Of course, we do not know the precise form of the geometric suppression, so we cannot say that it is precisely the Bekenstein assumption that is sufficient (or necessary), and hence we cannot deduce the $2\pi$ coefficient in the Bekenstein assumption from this argument, but it does suggest that at least for very large grireballs, something rather like the Bekenstein assumption might be necessary to avoid black holes' decaying too rapidly into grireballs.  This argument certainly does not rule out objects having entropy exceeding the Bekenstein assumption by large factors (see, e.g., \cite{Page:1982fj, Page:2000uq}, though I do not know of objects that violate the assumption {\it and} also have very large entropies), so long as the excess is not so large that when its exponential is multiplied by the very slow Hawking emission rate for zero-entropy objects of the same mass, it gives a total emission rate for all the different microstates of the grireballs that would cause observed astrophysical black holes to decay away in times less than their presumed ages.

One might wonder whether grireballs are related to black holes that might have firewalls.  So far I have discussed grireballs as separate objects, though as objects which might be emitted by black holes, or which black holes might transform into after very long times if the grireballs of the same mass as black holes are much larger in size.

One might assume that grireballs for a given mass are only slightly larger than black holes of the same mass, so that it would be easier for black holes to convert into them, and yet that the conversion might not be noticed in astrophysical observations, which typically only see the effects of the gravitational field beyond several Schwarzschild radii of the objects.  However, if the grireballs are not much larger than black holes but are highly hyper-entropic, with entropies much larger than black holes of the same mass, than not only would they violate the Bekenstein assumption (which might not be that noticeable if astronomical grireballs observationally look very much like black holes), but also what I have called the Hawking assumption, that $S \leq \pi R^2$.  This would seem more surprising, though if it is the Bousso bound \cite{Bousso:1999xy} that is the most compelling, that bound need not be violated if the grireballs do not have spacetime inside to be crossed by the null sheets in Bousso's covariant entropy conjecture.  Furthermore, if grireballs can indeed greatly exceed the Bekenstein assumption by huge entropy excesses, then they presumably would be rapidly emitted by significantly larger black holes.

Another possibility for grireballs, different from what was assumed above, is that do not form a one-parameter family, with both the radius $R$ and the entropy $S$ being given by functions of the mass $M$ alone, but that the mass and entropy are independent, with the radius being a function of both, $R = R(M,S)$.  Then, for example, one might imagine a scenario in which grireballs have $R(M,S) = S/(2\pi M)$ but only exist inside black holes for $S \leq 4 \pi M^2$ and hence then are not visible.  Only for $S > 4\pi M^2$ would grireballs be visible.

One might suppose, though it strains the imagination a bit, that the entropy $S$
that the grireball size depends on is the fine-grained von Neumann entropy
$S_\mathrm{vN}$, which is less than $4\pi M^2$ for black holes until they have
emitted enough radiation to become nearly maximally entangled with the radiation
and hence be classified as old black holes with fine-grained von Neumann entropy
very nearly the same as the coarse-grained thermodynamic entropy $S = 4\pi M^2$
for nonrotating black holes.  Then if such grireballs exist with radii
$R(M,S_\mathrm{vN}) = S_\mathrm{vN}/(2\pi M)$ for $S_\mathrm{vN} > 4 \pi M^2$,
once a black hole becomes old, it could become such a grireball and hence not merely have the possibility of a firewall just inside $R = 2M$, but a grireball boundary that is presumably a firewall at $R = S_\mathrm{vN}/(2\pi M) > 2M$.

That is, it seems conceivable that not only might old black holes develop firewalls at the Schwarzschild radius (which are hence not visible from the outside), but perhaps firewalls that expand outside the Schwarzschild radius and hence are visible.  Then it might be the case that even from the outside Bob could tell whether Alice is burning or fuzzing \cite{Chowdhury:2012vd}.

I have benefited from discussions with many colleagues over the years, and on the related subject of firewalls at the KITP Bits, Branes, and Black Holes program, though I have not yet subjected these current ideas to discussion.  As this paper was being finalized, a paper appeared \cite{Rama:2012fm} which discusses a `singularity cloud' which, depending on its properties and in particular whether it could be hyper-entropic, might be similar to some kind of grireball.  The final revisions of this paper were written at the Cook's Branch Nature Conservancy, where I have greatly appreciated the hospitality of the Mitchell family and of the George P. and Cynthia W. Mitchell Institute for Fundamental Physics and Astronomy of Texas A \& M University.  This work was supported in part by the Natural Sciences and Engineering Council of Canada.


\baselineskip 4pt

\begin{thebibliography}{99}

\bibitem{AMPS} 
  A.~Almheiri, D.~Marolf, J.~Polchinski and J.~Sully,
  ``Black Holes: Complementarity or Firewalls?,''
  arXiv:1207.3123 [hep-th].
  
\bibitem{Susskind:2012rm} 
  L.~Susskind,
  ``Singularities, Firewalls, and Complementarity,''
  arXiv:1208.3445 [hep-th]. 
  
\bibitem{Susskind:2012uw} 
  L.~Susskind,
  ``The Transfer of Entanglement: The Case for Firewalls,''
  arXiv:1210.2098 [hep-th]. 
  
\bibitem{Bousso:2012as} 
  R.~Bousso,
  ``Complementarity Is Not Enough,''
  arXiv:1207.5192 [hep-th].
  
\bibitem{Giveon:2012kp} 
  A.~Giveon and N.~Itzhaki,
  ``String Theory Versus Black Hole Complementarity,''
  arXiv:1208.3930 [hep-th].  
 
\bibitem{Nomura:2012sw} 
  Y.~Nomura, J.~Varela and S.~J.~Weinberg,
  ``Complementarity Endures: No Firewall for an Infalling Observer,''
  arXiv:1207.6626 [hep-th].  
  
\bibitem{Mathur:2012jk} 
  S.~D.~Mathur and D.~Turton,
  ``Comments on Black Holes I: The Possibility of Complementarity,''
  arXiv:1208.2005 [hep-th].  
  
\bibitem{Chowdhury:2012vd} 
  B.~D.~Chowdhury and A.~Puhm,
  ``Is Alice Burning or Fuzzing?,''
  arXiv:1208.2026 [hep-th].

\bibitem{Banks:2012nn} 
  T.~Banks and W.~Fischler,
  ``Holographic Space-Time Does Not Predict Firewalls,''
  arXiv:1208.4757 [hep-th].  

\bibitem{Ori:2012jx} 
  A.~Ori,
  ``Firewall or Smooth Horizon?,''
  arXiv:1208.6480 [gr-qc].    

\bibitem{Brustein:2012jn} 
  R.~Brustein,
  ``Origin of the Blackhole Information Paradox,''
  arXiv:1209.2686 [hep-th].

\bibitem{Hossenfelder:2012mr} 
  S.~Hossenfelder,
  ``Comment on the Black Hole Firewall,''
  arXiv:1210.5317 [gr-qc].

\bibitem{Nomura:2012cx} 
  Y.~Nomura, J.~Varela and S.~J.~Weinberg,
  ``Black Holes, Information, and Hilbert Space for Quantum Gravity,''
  arXiv:1210.6348 [hep-th].

\bibitem{Avery:2012tf} 
  S.~G.~Avery, B.~D.~Chowdhury and A.~Puhm,
  ``Unitarity and Fuzzball Complementarity: 'Alice Fuzzes but May not Even Know It!',''
  arXiv:1210.6996 [hep-th].

\bibitem{Larjo:2012jt} 
  K.~Larjo, D.~A.~Lowe and L.~Thorlacius,
  ``Black Holes without Firewalls,''
  arXiv:1211.4620 [hep-th].

\bibitem{Rama:2012fm} 
  S.~K.~Rama,
  ``Remarks on Black Hole Evolution a la Firewalls and Fuzzballs,''
  arXiv:1211.5645 [hep-th].
  
\bibitem{Bena:2012zi} 
  I.~Bena, A.~Puhm and B.~Vercnocke,
  ``Non-extremal Black Hole Microstates: Fuzzballs of Fire or Fuzzballs of Fuzz?,''
  arXiv:1208.3468 [hep-th].

\bibitem{Hwang:2012nn} 
  D.-i.~Hwang, B.-H.~Lee and D.-h.~Yeom,
  ``Is the Firewall Consistent?: Gedanken Experiments on Black Hole Complementarity and Firewall Proposal,''
  arXiv:1210.6733 [gr-qc].
  
\bibitem{Braunstein:2009my} 
  S.~L.~Braunstein, S.~Pirandola and K.~\.{Z}yczkowski,
  ``Entangled Black Holes as Ciphers of Hidden Information,''
  Physical Review Letters {\bf 110}, 101301S (2013)
  [arXiv:0907.1190 [quant-ph]].

\bibitem{Hayden:2007cs} 
  P.~Hayden and J.~Preskill,
  ``Black Holes as Mirrors: Quantum Information in Random Subsystems,''
  JHEP {\bf 0709}, 120 (2007)
  [arXiv:0708.4025 [hep-th]].

\bibitem{Boul}
  D.~G.~Boulware, 
  ``Quantum Field Theory in Schwarzschild and Rindler Spaces,''
  Phys.\ Rev.\ D{\bf 11}, 1404-1423 (1975).
  
\bibitem{Unruh}
  W.~G.~Unruh, 
  ``Notes on Black Hole Evaporation,''
  Phys.\ Rev.\ D{\bf 14}, 870-892 (1976).
  
\bibitem{HH}
  J.~B.~Hartle and S.~W.~Hawking,
  ``Path Integral Derivation of Black Hole Radiance,''
  Phys.\ Rev.\ D{\bf 13}, 2188-2203 (1976).
  
\bibitem{MS}
  C.~W.~Misner and D.~H.~Sharp,
  ``Relativistic Equations for Adiabatic, Spherically Symmetric Gravitational Collapse,''
  Phys.\ Rev.\ 136, B571-B576 (1964).

\bibitem{Bek}
  J.~D.~Bekenstein,  
  ``A Universal Upper Bound on the Entropy to Energy Ratio for Bounded Systems,''
  Phys.\ Rev.\ D{\bf 30}, 287-298 (1981).

\bibitem{Haw}
  S.~W.~Hawking, 
  ``Particle Creation by Black Holes,''
  Commun.\ Math.\ Phys.\ {\bf 43}, 199-200 (1975).

\bibitem{Page:1993df} 
  D.~N.~Page,
  ``Average Entropy of a Subsystem,''
  Phys.\ Rev.\ Lett.\  {\bf 71}, 1291 (1993)
  [gr-qc/9305007].
  
\bibitem{Page:1993wv} 
  D.~N.~Page,
  ``Information in Black Hole Radiation,''
  Phys.\ Rev.\ Lett.\  {\bf 71}, 3743 (1993)
  [hep-th/9306083].

\bibitem{RPP}
  J.~Beringer {\it et al.}  (Particle Data Group),
  ``Review of Particle Physics (RPP),''
  Phys.\ Rev.\ D {\bf 86}, 010001 (2012).

\bibitem{Page1}
  D.~N.~Page, 
  ``Particle Emission Rates from a Black Hole: Massless Particles from an Uncharged, Nonrotating Hole,''
  Phys.\ Rev.\ D{\bf 13}, 198-206 (1976).
  
\bibitem{Page2}
  D.~N.~Page, 
  ``Particle Emission Rates from a Black Hole. 2. Massless Particles from a Rotating Hole,''
  Phys.\ Rev.\ D{\bf 14}, 3260-3273 (1976).

\bibitem{Page3} 
  D.~N.~Page,
  ``Comment On 'Entropy Evaporated by a Black Hole',''
  Phys.\ Rev.\ Lett.\  {\bf 50}, 1013 (1983).

\bibitem{Page4}
D.~N.~Page,
  ``Hawking Radiation and Black Hole Thermodynamics,''
  New J.\ Phys.\  {\bf 7}, 203 (2005)
  [hep-th/0409024].

\bibitem{Page:1983ug} 
  D.~N.~Page,
  ``Comment on `Entropy Evaporated By A Black Hole',''
  Phys.\ Rev.\ Lett.\  {\bf 50}, 1013 (1983).

\bibitem{Page:2013dx} 
  D.~N.~Page,
  ``Time Dependence of Hawking Radiation Entropy,''
  arXiv:1301.4995 [hep-th].

\bibitem{Agnesi}
  D.~N.~Page, 
  ``Agnesi Weighting for the Measure Problem of Cosmology,''
  JCAP {\bf 1103}, 031 (2011).
  
\bibitem {abinitio}
  D.~N.~Page,
  ``Ab Initio Estimates of the Size of the Observable Universe,''
  JCAP {\bf 1109}, 037 (2011).

\bibitem{Hinshaw:2012fq} 
  G.~Hinshaw, D.~Larson, E.~Komatsu, D.~N.~Spergel, C.~L.~Bennett, J.~Dunkley, M.~R.~Nolta and M.~Halpern {\it et al.},
  ``Nine-Year Wilkinson Microwave Anisotropy Probe (WMAP) Observations: Cosmological Parameter Results,''
  arXiv:1212.5226 [astro-ph.CO].

\bibitem{Ade:2013xsa} 
  P.~A.~R.~Ade {\it et al.}  [Planck Collaboration],
  %``Planck 2013 results. I. Overview of products and scientific results,''
  arXiv:1303.5062 [astro-ph.CO].
  
\bibitem{Page:1982fj} 
  D.~N.~Page,
  ``Comment On A Universal Upper Bound On The Entropy To Energy Ratio For Bounded Systems,''
  Phys.\ Rev.\ D {\bf 26}, 947 (1982).

\bibitem{Page:2000uq} 
  D.~N.~Page,
  ``Defining Entropy Bounds,''
  JHEP {\bf 0810}, 007 (2008)
  [hep-th/0007238].  
  
\bibitem{Bousso:1999xy} 
  R.~Bousso,
  ``A Covariant Entropy Conjecture,''
  JHEP {\bf 9907}, 004 (1999)
  [hep-th/9905177].  
  
  
\end{thebibliography}
\end{document}